\documentclass[12pt]{iopart} 
\usepackage{graphicx}
\DeclareSymbolFont{AMSb}{U}{msb}{m}{n}
\DeclareMathSymbol{\R}{\mathalpha}{AMSb}{"52}
\begin{document}

\title{Critical strength of attractive central potentials}
\author{Fabian Brau\footnote[1]{E-Mail: fabian.brau@umh.ac.be}}
\address{Groupe de Physique Nucl\'eaire Th\'eorique, Universit\'e de Mons-Hainaut, Acad\'emie Universitaire Wallonie-Bruxelles, B-7000 Mons, Belgique}
\author{Monique Lassaut\footnote[2]{E-Mail: lassaut@ipno.in2p3.fr}}
\address{Groupe de Physique Th\'eorique, Institut de Physique Nucl\'eaire, F-91406 Orsay CEDEX, France}
\date{\today}

\begin{abstract}
We obtain several sequences of necessary and sufficient conditions for the existence of bound states applicable to attractive (purely negative) central potentials. These conditions yields several sequences of upper and lower limits on the critical value, $g_{\rm{c}}^{(\ell)}$, of the coupling constant (strength), $g$, of the potential, $V(r)=-g v(r)$, for which a first $\ell$-wave bound state appears, which converges to the exact critical value.
\end{abstract}

\maketitle

\section{Introduction}
\label{sec1}

Since the pioneer works of Jost and Pais in 1951 \cite{jost51} and Bargmann in 1952 \cite{barg52}, the determination of upper and lower limits on the number of bound states of a given potential, having spherical symmetry $V(r)$, in the framework of non relativistic quantum mechanics is still of interest. A fairly large number of results of this kind can be found in the literature for the Schr\"odinger equation (see for example \cite{sc61,ca65a,ca65b,ca65c,ca65d,ch68,gl76,si76,ma77,li80,ch95a,ch96,bl96,la97,br03a,br03b} and  for results applicable to one and two dimension spaces see for example \cite{new62,gla78,new83,ak98,cha03})

An important theorem for classifying these results was found by Chadan \cite{ch68} and gives the asymptotic behavior of the number of $\ell$-wave bound states as the strength, $g$, of the central potential $V(r)=g v(r)$ goes to infinity:
\begin{equation}
\label{eq0}
N_{\ell} \approx \frac{g^{1/2}}{\pi}\int_0^{\infty}dr\, v^-(r)^{1/2}\quad {\rm as}\quad g\rightarrow \infty,
\end{equation}
where the symbol $\approx$ means asymptotic equality, $V^-(r)=g v^-(r)$ and $v^-(r)=\max(0,-v(r))$ (see also Ref. \cite{ma72} for a generalization of relation (\ref{eq0})). This result implies that any upper and lower limit on $N_{\ell}$ which could yield cogent results should behave asymptotically as $g^{1/2}$. More importantly, the relation (\ref{eq0}) gives the functional of the potential, that is to say, the coefficient in front of $g^{1/2}$ that appears in the asymptotic behavior. Upper and lower limits on the number of $\ell$-wave bound states featuring the correct $g^{1/2}$ dependency was first obtained in Ref. \cite{ca65c}. Upper and lower limits on $N_{\ell}$ featuring the correct asymptotic behavior (\ref{eq0}) was first derived in Refs. \cite{br03a,br03b}. In practice, the asymptotic regime is reached very quickly when the strength of the potential is large enough to bind two or three bound states.

The situation is completely different when one consider the transition between zero and one bound state. In contrast to one and two dimension cases where any attractive potential, satisfying adequate integrability conditions, has at least one  bound state, in the three dimensional case, the potential acquires a bound state only if it is attractive (negative) enough.
Thus, for central potential for example, there exists a ``critical" value, $g_{\rm{c}}^{(\ell)}$, of the coupling constant (strength), $g$, of the potential, $V(r)=g v(r)$, for which a first $\ell$-wave bound state appears. The determination of this critical value requires to solve the zero energy Schr\"odinger equation \cite{sc61,bip61,sim71}. To circumvent the exact calculation of the Jost function at zero energy, upper and lower bounds are very helpful. From now on, use will be made of the standard quantum-mechanical units $\hbar=2\mu=1$ where $\mu$ is the reduced mass of the particles.

In 1976, Glaser {\it et al.} have obtained a strong necessary condition for the existence of a $\ell$-wave bound state in an arbitrary central potential in three dimensions \cite{gl76}:
\begin{equation}
\label{eq1}
(\forall p \geq 1) \quad  \frac{(p-1)^{p-1}\,\Gamma (2p)}{(2\ell+1)^{2p-1}\,p^p\, \Gamma^2(p)} \int_{0}^{\infty}\frac{dr}{r}\,\left[r^2\, V^-(r)\right]^p \geq 1.  
\end{equation}
This relation yields a lower limit on the critical value $g_{\rm{c}}^{(\ell)}$, by making a minimization over $p \geq 1$, which was shown to be very accurate (see for example \cite{gl76,la97,bra03}).

Other necessary conditions for the existence of bound states can be found in the literature (see for example \cite{ca65d,ma77,bra03} and for reviews see \cite{si76,br03a,br03b}), but in general, the relation (\ref{eq1}) yields the strongest restriction on $g_{\rm{c}}^{(\ell)}$ (in some cases, the relations obtained in Ref. \cite{bra03} can however be better).

Sufficient conditions for existence of bound states, yielding upper limits on the critical value of the strength of the potential, are scarcer. Two sufficient conditions for the existence of at least one bound state with angular momentum $\ell$ have been found by Calogero in 1965 \cite{ca65b,ca65c}
\begin{equation}
\label{eq2}
\fl (\forall R >0) \quad \int_0^R dr\, r\, |V(r)|\, (r/R)^{2\ell+1}+\int_R^{\infty}dr\, r\, |V(r)|\,(r/R)^{-(2\ell+1)}>2\ell+1,
\end{equation}
and
\begin{equation}
\label{eq3}
(\forall R >0) \quad R \int_0^{\infty} dr\, |V(r)|\left[(r/R)^{2\ell}+(r/R)^{-2\ell}\,R^2|V(r)|\right]^{-1}>1.
\end{equation}
These two conditions apply provided the potential is nowhere positive, $V(r)=-|V(r)|$. The most stringent conditions obtain by minimizing the left-hand sides of (\ref{eq2}) and (\ref{eq3}) over all positive values of $R$. Some other sufficient conditions for the existence of bound states can been found in the literature (see for example \cite{la97,bra03} and for reviews see \cite{si76,br03a,br03b}). A sufficient condition which does not require the spherical symmetry for the potential $V$ was proposed in 1980 by Chadan \cite{cha79} (see also \cite{cha80}). When the potential is central and purely attractive, the inequality:
\begin{equation}
{\rm Tr} K^{(2)}  \geq  {\rm Tr} K^{(1)},
\label{chad}
\end{equation}
where 
\begin{eqnarray}
K^{(1)}(r,r') & = & \inf(r,r')^{\ell+1} \ \sup(r,r')^{-\ell} \ V(r') \nonumber\\
K^{(n)}(r,r') & = &\int_0^{\infty} ds \ K^{(1)}(r,s) \ K^{(n-1)}(s,r'),
\end{eqnarray}
implies the existence of at least a bound state for the potential $V(r)$. In the case where $V(r)$ has some changes of sign, the condition (\ref{chad}), is replaced by
\begin{equation}
{\rm Tr} K^{(4)}  \geq  {\rm Tr} K^{(2)},
\label{chad1}
\end{equation}
which implies that one of the potentials $\pm V(r)$ has at least one bound state.

Recently, an upper limit on the critical strength has been found, originating from a variational technique \cite{bra03d}:
\begin{equation}
\label{eq4}
\fl g_{\rm{c}}^{(\ell)}\leq \lambda \,\int_0^{\infty} dx\, F(2p-1;x) \left[\int_0^{\infty} dy\, F(p;y) y^{-\lambda}\int_0^y dz\, F(p;z) z^{\lambda}\right]^{-1},
\end{equation}
with $F(q;x)=x^q\, v(x)^{(q+1)/2}$, $v(x)\ge 0$, $\lambda=\ell+1/2$ and $q>0$ which was found to be very accurate. Clearly more accurate upper limits could be obtained but depending strongly on the choice of the trial wave function.

The present article follows a different scheme. To circumvent the  difficulty to guess a trial (strongly potential dependent) wave function for variational methods we propose upper and lower limits originating  from iterative procedures, 
all designed to converge towards the exact result. The methods proposed by Chadan  enter this category of iterative convergent procedures. Also, Lassaut and Lombard \cite{la97} worked in this sense some years ago, but the procedure was constrained by the condition $g_{\rm{c}}^{(\ell)} \int_0^{\infty} dr \, r |v(r)| <2 (2\ell + 1)$, which is not verified when the convexity of the potential is too high or when the angular momentum $\ell$ increases. We thus propose in this article sequences of upper and lower limits of the critical value, $g_{\rm{c}}^{(\ell)}$, which converge towards the critical coupling constant without any restriction on the possible values of $g_{\rm{c}}^{(\ell)}$. The advantage of our procedure, based upon the Riesz theorem \cite{schw,riesz} in what concerns the upper limits, is that there is no need to start from a conveniently chosen wave function. Indeed, the improvements are simple for these sequences: one just need to calculate the next order. Note that the basic idea of sequences of lower limits for $g_{\rm{c}}^{(\ell)}$ have already been explored in Ref. \cite{bra03}.

The paper is organized as follows. In section \ref{sec2} we derive the upper limits on the critical value $g_{\rm{c}}^{(\ell)}$. In section \ref{sec3} the algorithm for generating both lower and upper limits is discussed. In section \ref{sec4} our proposal of upper and lower bounds are tested against the exact values for common potentials. 
Our conclusions are presented in section \ref{sec5}.

\section{Upper limits on the critical strength}
\label{sec2}

From now on, we assume that $V(r)$ is locally integrable and such that 
\begin{equation}
\label{intv}
\int_0^{\infty} dr\, r\, |V(r)| < \infty, 
\end{equation}
remembering that we consider purely attractive potentials namely satisfying $V(r) \leq 0$. Following Birman and Schwinger \cite{sc61,bip61,sim71} the critical values of the strength of the potential correspond to the occurrence of an eigenstate with a vanishing energy. In this paper we consider the zero energy Schr\"odinger equation that we write into the form of an integral equation incorporating the boundary conditions
\begin{equation}
\label{eq5}
u_{\ell}(r)=-\int_0^{\infty}dr'\, g_{\ell}(r,r')\, V(r')\, u_{\ell}(r'),
\end{equation}
where $g_{\ell}(r,r')$ is the Green function of the kinetic energy operator and is explicitly given by
\begin{equation}
\label{eq6}
g_{\ell}(r,r')=\frac{1}{2\ell+1} r_<^{\ell+1}\, r_>^{-\ell},
\end{equation}
where $r_<=\inf(r,r')$ and $r_>=\sup(r,r')$. An important technical difficulty appears if the potential possesses some changes of sign (see relation (\ref{eq7}) below). This is overcome for the derivation of necessary conditions, or of upper bounds on the number of bound states, by replacing the potential by its negative part $V(r)\rightarrow V^-(r)=\max(0,-V(r))$. Indeed, the potential $V^-(r)$ is more attractive than $V(r)$ and thus a necessary condition for existence of bound states in $V^-(r)$ is certainly a valid necessary condition for $V(r)$. This procedure can no longer be used to obtain sufficient conditions. For this reason we consider potentials that are nowhere positive, $V(r)=-g v(r)$, with $v(r)\geq 0$.

To obtain a symmetrical kernel we now introduce the function $\psi_{\ell}(r)$ as follow
\begin{equation}
\label{eq7}
\psi_{\ell}(r)=\sqrt{v(r)} \, u_{\ell}(r).
\end{equation}
Equation (\ref{eq5}) becomes
\begin{equation}
\label{eq8}
\psi_{\ell}(r)=g\int_0^{\infty}dr'\, K_{\ell}(r,r')\, \psi_{\ell}(r'),
\end{equation}
where the symmetric kernel $K_{\ell}(r,r')$ is given by
\begin{equation}
\label{eq9}
K_{\ell}(r,r')=\sqrt{v(r)}\, g_{\ell}(r,r')\,\sqrt{v(r')}.
\end{equation}
The relation (\ref{eq8}) is thus an eigenvalue problem with a symmetric kernel and, for each value of $\ell$, the smallest characteristic numbers are just the critical values $g_{\rm{c}}^{(\ell)}$. The higher characteristic numbers correspond to the critical values of the strength for which a second, a third, ..., $\ell$-wave bound state appears. The kernel (\ref{eq9}) acting on the Hilbert space $L^2(\R)$ is an Hilbert-Schmidt kernel \cite{chad86} for the class of potentials defined by (\ref{intv}), i.e. satisfies the inequality
\begin{equation}
\int_0^{\infty} \int_0^{\infty} dx\, dy\, K_{\ell}(x,y)K_{\ell}(x,y) < \infty .
\end{equation}
Consequently the eigenvalue problem (\ref{eq8}) always possesses at least one characteristic number \cite{tri1} (in general, 
this problem has an infinity of characteristic numbers). 

We propose now to solve the eigenvalue problem (\ref{eq8}) using iterative methods. 

\subsection{Iterative power method}
\label{sec2.1}

Let us write, for simplicity, the relation (\ref{eq8}) under the form
\begin{equation}
\label{eq10}
\psi_{\ell}=g\, {\cal K}_{\ell}\, \psi_{\ell},
\end{equation}
where ${\cal K}_{\ell}$ denotes the symmetric linear operator, operating on the Hilbert space $L^2(\R)$, which is in this paper the integral operator generated by the so-called Birman-Schwinger \cite{sc61,bip61} kernel $K_{\ell}$, Eq. (\ref{eq9}).   

Since the kernel $K_{\ell}(r,r')$ is Hilbert-Schmidt, ${\cal K}_{\ell}$ is a compact operator \cite{newt}. As ${\cal K}_{\ell}$ is symmetric the Riesz theorem applies \cite{schw,riesz}. For {\it each} value of the angular momentum $\ell$, the set of eigenvalues $1/g_p$, $1 \leq p$, (which in the present case are all positive) can be ordered according to a sequence tending to zero, $1/g_1 \geq 1/g_2 \geq \ldots \geq 1/g_p\geq \ldots \geq 0$. There exists an orthonormal basis in $L^2(\R)$, labeled $\varphi_p(r)$, $p \geq 1$, each $\varphi_p(r)$ being associated to $1/g_p$, and for each function $\phi_{\ell}(r) \in L^2(\R)$ 
\begin{equation}
\label{riesz}
{\cal K}_{\ell} \, \phi_{\ell} =\sum_{p=1}^{\infty} \langle  {\cal K}_{\ell} \, \phi_{\ell} | \varphi_p \rangle \varphi_p = \sum_{p=1}^{\infty} \langle  \phi_{\ell} |   {\cal K}_{\ell} \, \varphi_p \rangle \varphi_p = \sum_{p=1}^{\infty} \frac{1}{g_p} \,  \langle \phi_{\ell} |\varphi_p \rangle \, \varphi_p,
\end{equation}
where the symbol $\langle f | g \rangle $ denotes the scalar product $\int_0^{\infty} dr \, f(r) \,  g(r)$. For the sake of simplicity we have dropped the indices $(\ell)$ which should appear on $g_p$ and on $\varphi_p(r)$.

The positivity of the eigenvalues originates \cite{riesz} from the fact that ${\cal K}_{\ell}$ is positive  i.e. 
\begin{equation}
(\forall\, \phi_{\ell} \in L^2(\R)) \quad\quad \langle \phi_{\ell} |{\cal K}_{\ell} \, \phi_{\ell}\rangle \geq 0 .
\end{equation}

In the present case, where the potential has the spherical symmetry, there is no degeneracy and we have strict inequalities for the eigenvalues
\begin{equation}
\label{ineq}
\frac{1}{g_1} > \frac{1}{g_2} > \ldots >  \frac{1}{g_p} > \ldots > 0.
\end{equation}
This is due to the fact that the eigenstates are solutions of a linear second order equation with constraints at the origin and infinity.

Now, we introduce the iterated kernel $K_{\ell}^{(n)}(s,t)$ of $K_{\ell}(s,t)$
\begin{equation}
K_{\ell}^{(n)}(s,t)=\int_0^{\infty} du \, K_{\ell}(s,u)\, K_{\ell}^{(n-1)}(u,t) \quad n \geq 2,
\end{equation}
with  
\begin{equation}
K_{\ell}^{(1)}(s,t)=K_{\ell}(s,t). 
\end{equation}
We can then compute the scalar product between $\phi_{\ell}$ and ${\cal K}_{\ell}^{(n)}\, \phi_{\ell}$, and find
\begin{equation}
\langle \phi_{\ell} |{\cal K}_{\ell}^{(n)} \phi_{\ell} \rangle =\langle  {\cal K}_{\ell}^{(n)} \phi_{\ell} | \phi_{\ell} \rangle= \sum_{p=1}^{\infty} \frac{1}{g_p^n} \, \langle \phi_{\ell}|\varphi_p \rangle^2. 
\label{ries}
\end{equation}
Therefore, we obtain the following convexity-type relation
\begin{eqnarray}
\fl \langle \phi_{\ell} | {\cal K}_{\ell}^{(n+1)} \phi_{\ell} \rangle\, \langle \phi_{\ell} |{\cal K}_{\ell}^{(n-1)} \phi_{\ell} \rangle - \langle \phi_{\ell} |{\cal K}_{\ell}^{(n)} \phi_{\ell} \rangle^2&=& \sum_{p<q=1}^{\infty} \frac{1}{(g_p g_q)^{n-1}} \left( \frac{1}{g_p}-\frac{1}{g_q} \right)^2 \nonumber \times \\ &\times& \langle \phi_{\ell}|\varphi_p \rangle^2 \, \langle \phi_{\ell}|\varphi_q \rangle^2, \nonumber \\ 
&\geq& 0.
\end{eqnarray}
Since the left hand side of this latter equality is positive we conclude that the sequence 
\begin{equation}
\label{delt}
n \mapsto \delta_n=\frac{\langle \phi_{\ell} |{\cal K}_{\ell}^{(n+1)} \phi_{\ell} \rangle}{
\langle \phi_{\ell} |{\cal K}_{\ell}^{(n)} \phi_{\ell} \rangle} \quad n \geq 1,
\end{equation}
with
\begin{equation}
\delta_0=\frac{\langle \phi_{\ell} |{\cal K}_{\ell} \phi_{\ell} \rangle}{\langle \phi_{\ell} |\phi_{\ell} \rangle},
\end{equation}
is always {\it increasing}. 

On the other hand, taking into account (\ref{ries}) and (\ref{delt}) we have
\begin{equation}
\delta_n= \frac{1}{g_1} \frac{\sum_{p=1}^{\infty} (\frac{g_1}{g_p})^{n+1} \, \langle \phi_{\ell}|\varphi_p \rangle^2 }{ \sum_{p=1}^{\infty} (\frac{g_1}{g_p})^n \, \langle \phi_{\ell}|\varphi_p \rangle^2 } .
\end{equation}
Due to the strict inequalities (\ref{ineq}) and the Parseval's formula 
\begin{equation}
\vert \vert \phi_{\ell} \vert \vert_2^2=\sum_{p=1}^{\infty}\langle \phi_{\ell} |\varphi_p \rangle^2,
\label{pars}
\end{equation}
the sequence $\delta_n$ converges to $1/g_1$, where  $g_1=g_{\rm{c}}^{(\ell)}$ is the lowest critical value of the coupling constant of the potential, except when $ \langle \phi_{\ell}|\varphi_1 \rangle $ is zero. This procedure, known in the literature as the iterated power method, yields the maximal eigenvalue of the problem considered, i.e. in the present case to the lowest critical value $g_{\rm{c}}^{(\ell)}=g_1$. Any starting positive (non zero) squared integrable function $\phi_{\ell}(r)$ is appropriated. Indeed the function $\varphi_1(r)$ has no node which insures that the scalar product $ \langle \phi_{\ell}|\varphi_1 \rangle$ is not zero. (In our numerical studies, presented in section \ref{sec4} use is made of the choice $\phi_{\ell}(r)=r^{\ell+1}\sqrt{v(r)}$). Moreover since $n \mapsto \delta_n$ is increasing we get upper bounds for $g_{\rm{c}}^{(\ell)}$ namely $g_{\rm{c}}^{(\ell)} < \ldots  < 1/\delta_p < \ldots < 1/\delta_2 < 1/\delta_1$.

There exists other iterative methods in the literature and we discuss briefly two variants of them in the next two sections.

\subsection{Kellogg's method}
\label{sec2.2}

In this section, we consider the method proposed by Kellogg for the compact operator ${\cal K}_{\ell}$ \cite{kel,riesz}. 
We construct the following sequence of functions
\begin{equation}
\label{eq10z}
\phi_{\ell}^{(n+1)}(r)=\int_0^{\infty} dr' K_{\ell}(r,r')\, \phi_{\ell}^{(n)}(r'),
\end{equation}
where $K_{\ell}(r,r')$ is given by (\ref{eq9}). This latter relation is schematically written as:
\numparts
\label{eq11}
\begin{equation}
\label{eq11a}
\phi_{\ell}^{(n+1)}={\cal K}_{\ell}\, \phi_{\ell}^{(n)}.
\end{equation}
For non zero $\phi_{\ell}^{(0)}(r)$, Kellogg considers the following sequence of numbers
\begin{equation}
\label{eq11b}
\gamma_{n+1}=\frac{||\phi_{\ell}^{(n)}||_2}{||\phi_{\ell}^{(n+1)}||_2} \quad n\geq 1,
\end{equation}
\endnumparts
where $||\phi_{\ell}^{(n)}||_2$ is the $L^2$ norm of the function $\phi_{\ell}^{(n)}(r)$. 

We keep our conventions namely $\varphi_1(r)$, $\varphi_2(r)$, ..., still denote  the  eigenfunctions of the problem (\ref{eq10}) and $g_1 <g_2 < \ldots$  are the corresponding characteristic numbers. 

Suppose that $\phi_{\ell}^{(0)}(r)$ is orthogonal to the functions $\varphi_1(r)$, $\varphi_2(r)$, ..., $\varphi_{k-1}(r)$ but not orthogonal to the function $\varphi_k(r)$. Then the sequence $\gamma_n$ converges toward $g_k$ with the property that $g_k \leq \gamma_n$. Moreover the sequence of functions $\phi_{\ell}^{(n)}(r)/||\phi_{\ell}^{(n)}||_2$ converges to $\varphi_k(r)$ in $L^2(\R)$ \cite{riesz}. Consequently there exists a subsequence of $n \mapsto \phi_{\ell}^{(n)}(r)/||\phi_{\ell}^{(n)}||_2$ which converges almost everywhere to $\varphi_k(r)$.

The convergence of $\gamma_n$ is illustrated simply here, where we are interested by the smallest characteristic number
$g_{\rm{c}}^{(\ell)}$. Still we choose a positive (non zero) squared integrable function $\phi_{\ell}^{(0)}(r)$,
which, according to our previous discussion, is not orthogonal to $\varphi_1(r)$. The sequence of numbers $\gamma_n$ provides us upper bounds for $g_{\rm{c}}^{(\ell)}$ because this sequence is {\it decreasing} and converges to $g_{\rm{c}}^{(\ell)}$. The monotony of the sequence $n \mapsto \gamma_n$ is  related to the following equation, derived from (\ref{ries}),
\begin{equation}
\label{riesz2}
\vert \vert \phi_{\ell}^{(n)} \vert \vert_2^2=\langle {\cal K}_{\ell}^{(n)}\phi^{(0)}_{\ell}|{\cal K}_{\ell}^{(n)} \phi^{(0)}_{\ell} \rangle = \sum_{p=1}^{\infty} \frac{1}{g_p^{2n}} \, \langle \phi^{(0)}_{\ell}|\varphi_p \rangle^2 ,
\end{equation}
which leads to the inequality 
\begin{eqnarray}
\fl (\gamma_{n+2}^2-\gamma_{n+1}^2) ||\phi_{\ell}^{(n+2)}||_2^2 \ ||\phi_{\ell}^{(n+1)}||_2^2 &=& - \sum_{p<q=1}^{\infty}  \frac{1}{(g_p g_q)^{2n}} \left( \frac{1}{g_p^2}-\frac{1}{g_q^2} \right)^2\langle \phi^{(0)}_{\ell}| \varphi_p \rangle^2  \, \langle \phi^{(0)}_{\ell}| \varphi_q \rangle^2, \nonumber \\ &\leq& 0 \ .
\end{eqnarray}
The positivity of $\gamma_n$ asserts $\gamma_{n+2} \leq \gamma_{n+1}$. 

On the other hand we can check easily that $\gamma_n$ converges to $g_{\rm{c}}^{(\ell)}$ for $n$ going to infinity. Indeed, 
taking into account definitions (\ref{eq11b}) and (\ref{riesz2}) we have
\begin{equation}
\gamma_n^2= g_1^2 \  \frac{\sum_{p=1}^{\infty} (\frac{g_1}{g_p})^{2 n} \, 
\langle \phi_{\ell}^{(0)}|\varphi_p \rangle^2 }{ \sum_{p=1}^{\infty} 
(\frac{g_1}{g_p})^{2 n+2} \, \langle \phi_{\ell}^{(0)}|\varphi_p \rangle^2 },
\end{equation}
which, using (\ref{ineq}) and (\ref{pars}), shows clearly the convergence.

In the section \ref{sec4}, use will be made of the iterative procedure (\ref{eq10z}) but for the functions $u_{\ell}^{(n)}(r)$, defined by $u_{\ell}^{(n)}(r)=\phi_{\ell}^{(n)}(r)/\sqrt{v(r)}$, namely 
\begin{eqnarray}
\label{eq13}
\fl u_{\ell}^{(n+1)}(r)=\int_0^{\infty} dr'\, v(r') g_{\ell}(r,r') u_{\ell}^{(n)}(r'), \nonumber \\
\lo= \frac{r^{-\ell}}{2\ell+1}\int_0^r dr'\, r'^{\ell+1} v(r') u_{\ell}^{(n)}(r')+ \frac{r^{\ell+1}}{2\ell+1}\int_r^{\infty} dr'\,  r'^{-\ell} v(r') u_{\ell}^ {(n)}(r'),
\end{eqnarray}  
with $u_{\ell}^{(0)}(r)=r^{\ell+1}$. The sequence $\gamma_n$ is given by 
\begin{equation}
\label{eq14}
\gamma_{n+1}=\sqrt{\frac{\int_0^{\infty}dr\, v(r)\, [u_{\ell}^{(n)}(r)]^2}{
\int_0^{\infty}dr\, v(r)\, [u_{\ell}^{(n+1)}(r)]^2 } }.
\end{equation}

\subsection{Kolom\'y's method}
\label{sec2.3}

The second variant of iterative methods was proposed by Kolom\'y \cite{kol60} who constructed the sequence of functions
\numparts
\label{eq12}
\begin{equation}
\label{eq12a}
\phi_{\ell}^{(n+1)}=\frac{\langle {\cal K}_{\ell}\,  \phi_{\ell}^{(n)} |\phi_{\ell}^{(n)} \rangle}{||{\cal K}_{\ell}\, \phi_{\ell}^{(n)}||_2^2}\,{\cal K}_{\ell}\, \phi_{\ell}^{(n)},
\end{equation}
and consider the sequence of numbers
\begin{equation}
\label{eq12b}
\beta_{n+1}=\frac{\langle {\cal K}_{\ell}\, \phi_{\ell}^{(n)} | \phi_{\ell}^{(n)}  \rangle}{||{\cal K}_{\ell}\, \phi_{\ell}^{(n)}||_2^2}.
\end{equation}
\endnumparts
The factor $(\langle {\cal K}_{\ell}\, \phi_{\ell}^{(n)} |\phi_{\ell}^{(n)} \rangle)/(||{\cal K}_{\ell}\, \phi_{\ell}^{(n)}||_2^2)$ in (\ref{eq12a}) is useless as far as the series of number (\ref{eq12b}) is concerned. In this paper, where our interest is focused to the lowest eigenvalue of the problem, we simply drop the normalization factor in (\ref{eq12a}), namely we consider
\begin{equation}
\label{eq13ba}
\phi_{\ell}^{(n+1)}={\cal K}_{\ell}\, \phi_{\ell}^{(n)},
\end{equation}
coupled with (\ref{eq12b}).

When the starting trial function $\phi_{\ell}^{(0)}(r)$ is a positive (non zero) squared integrable function, it is  not orthogonal to $\varphi_1(r)$. Then the sequence $\beta_n$ converge to $g_1$ with $g_1 \leq \beta_n$. We still obtain a sequence of upper bounds for $g_{\rm{c}}^{(\ell)}$. Indeed the sequence $n \mapsto \beta_n$ is {\it decreasing} since  
\begin{eqnarray}
\fl (\beta_{n+2}-\beta_{n+1}) ||{\cal K}_{\ell}\, \phi_{\ell}^{(n+1)}||_2^2\ ||{\cal K}_{\ell}\, \phi_{\ell}^{(n)}||_2^2 &=& -\sum_{p<q=1}^{\infty} \frac{1}{(g_p  g_q)^{2n-1}} \, \left( \frac{1}{g_p}-\frac{1}{g_q} \right)^2 \, \left( \frac{1}{g_p}+\frac{1}{g_q} \right) \times \nonumber \\ &\times& \langle \phi^{(0)}_{\ell}| \varphi_p \rangle^2  \, \langle \phi^{(0)}_{\ell}| \varphi_q\rangle^2 \nonumber \\
&\leq& 0.
\end{eqnarray}

On the other hand we can check easily that $\beta_n$ converges to $g_1=g_{\rm{c}}^{(\ell)}$ for $n$ going to infinity. Indeed, taking into account definition (\ref{eq12b}) and (\ref{riesz}), we have
\begin{equation}
\beta_{n+1}= g_1 \  \frac{\sum_{p=1}^{\infty} (\frac{g_1}{g_p})^{2 n+1} \,\langle \phi_{\ell}^{(0)}|\varphi_p \rangle^2 }
{\sum_{p=1}^{\infty} (\frac{g_1}{g_p})^{2 n+2} \,\langle \phi_{\ell}^{(0)}|\varphi_p \rangle^2 },
\end{equation}
which, using (\ref{ineq}) and (\ref{pars}), shows clearly the convergence.

For the sake of convenience, we rewrite the iterative procedure (\ref{eq13ba}) in terms of the function $u_{\ell}^{(n)}(r)$, defined by $u_{\ell}^{(n)}(r)=\phi_{\ell}^{(n)}(r)/\sqrt{v(r)}$. Using (\ref{eq12b}), the sequence $\beta_n$ reads
\begin{equation}
\label{eq16}
\beta_{n+1}=\frac{\int_0^{\infty}dr\, v(r)\, u_{\ell}^{(n)}(r) \,u_{\ell}^{(n+1)}(r)}{\int_0^{\infty}dr\, v(r)\, [u_{\ell}^{(n+1)}(r)]^2 }.
\end{equation}
As we show in section \ref{sec4}, both iterative procedures depicted in \ref{sec2.2} and \ref{sec2.3} converge very rapidly, 
when use is made of the initial function $\phi_{\ell}^{(0)}(r)=r^{\ell+1}\sqrt{v(r)}$. However to decrease the number of iterations, more flexible function $\phi_{\ell}^{(0)}(r)$, obeying to the desired conditions, positivity and squared integrability, could be very helpful. This is explored in the next section.

\subsection{Combination of iterative and variational methods}
\label{sec2.4}

The variational method we propose is based upon the theorem \cite{riesz1,tri} which states that for a symmetric compact operator, 
\begin{equation}
\label{eq16b}
\sup_{\psi} [\langle {\cal K}_{\ell}\, \psi|\psi\rangle]=\frac{1}{\vert g_1\vert},
\end{equation}
under the constraint that $\psi$ is squared integrable and normalized to unity ($||\psi||_2=1$). Since in this article the kernel we consider is positive we have $\vert g_1 \vert =g_1$. The maximal value of (\ref{eq16b}) is reached for $\psi(r)=\varphi_1(r)$, the eigenfunction associated to the smallest eigenvalue $g_1= g_{\rm{c}}^{(\ell)}$. Consequently, for any function $\psi(r)$, normalized to unity, we obtain the following upper limit
\begin{equation}
\label{eq16c}
g_{\rm{c}}^{(\ell)}\leq \langle {\cal K}_{\ell}\, \psi|\psi\rangle^{-1}.
\end{equation}

This method was used in Ref. \cite{bra03d} to obtain the upper limit (\ref{eq4}) using the trial function
\begin{equation}
\label{eq16d}
\psi(r)=A\left[r^{2p-1} v(r)^p\right]^{1/2},\quad p>0.
\end{equation}

The degree of flexibility increases with  
\begin{equation}
\label{eq16e}
\psi(r)=A\left[r^p v(r)^q\right]^{1/2},
\end{equation}
where still $(p,q)$ are varied in the domain ensuring that $\psi(r) \in L^2(\R)$ and $A$ is a normalization factor designed to have $||\psi||_2=1$. When we combine the iterative methods together with the variational one, we construct a sequence of functions which automatically converges in $L^2(\R)$ to the exact zero energy wave function. This means that the variational method (\ref{eq16b}) can be used with all iterated $\psi^{(i)}={\cal K}_{\ell}^{(i)} \psi$, $i=1,2,\ldots$, of $\psi(r)$, Eq. (\ref{eq16e}).

These possibilities are tested in section \ref{sec4} and shown to greatly improve the convergence towards the exact result.

\section{Upper and lower limits on the critical strength}
\label{sec3}

Now, let us introduce another method which provide both lower and upper limits of the critical value required. Note that this method has some link with the iterative power method discussed above.

In this section we restrict to the S-wave case since the determination of the critical value for the potential $V(r)$ in the $\ell$-wave is equivalent to the determination of the critical value for the potential 
\begin{equation}
\label{W}
W_{\ell}(r) = \frac{1}{(2 \ell+1)^2} \frac{V(r^{1/(2 \ell+1)})}{r^{4 \ell/(2\ell+1)}}
\end{equation}
in the S-wave \cite{la97}.

{\bf Theorem} {\it Let $V(r)$ be a central potential with $V(r)=-g v(r), v(r) \geq  0$. Let the functions $\psi_n(r)$ 
defined by the following recurrence relation}
\begin{equation}
\psi_n(r)=\int_0^{\infty}dr'\, g(r,r')\, v(r') \ \psi_{n-1}(r') \quad n \geq 1
\label{psin}
\end{equation}
{\it with $\psi_0(r)=r$ and $g(r,r')=\inf(r,r')$.
Let $\alpha_n$ and $\omega_n$ be two sequences defined as follows,}
\begin{equation}
\label{alphan}
\alpha_n=\lim_{r\rightarrow 0}\frac{\psi_{n+1}(r)}{\psi_{n}(r)}=
\frac{\psi'_{n+1}(0)}{\psi'_{n}(0)} = 
\frac{\int_0^{\infty}dr\, v(r)\, \psi_{n}(r)}{\int_0^{\infty}dr\, 
v(r)\, \psi_{n-1}(r)}  \quad n \geq 1,
\end{equation}
where the prime denotes the derivative with respect to $r$ and
\begin{equation}
\label{bet}
\omega_n=\lim_{r\rightarrow \infty}\frac{\psi_{n+1}(r)}{\psi_{n}(r)}=
\frac{\int_0^{\infty}dr\, r\,v(r)\, \psi_{n}(r)}{\int_0^{\infty}dr\, 
r\,v(r)\, \psi_{n-1}(r)}  \quad n \geq 1.
\end{equation}
{\it Then, the two sequences $\alpha_n$ and $\omega_n$ converges to $1/g^{(0)}_{\rm{c}}$, $g^{(0)}_{\rm{c}}$ being the critical value of the strength of the potential $V(r)$ for which a first S-wave bound state appears. The $\alpha_n$ sequence is decreasing while the $\omega_n$ sequence is increasing.} 

The proof that $\omega_n$ is {\it increasing} and converges towards the desired value is rather simple. The change of function $\phi_{0}^{(n)}(r)=\sqrt{v(r)}\, \psi_n(r)$ in the iterative process (\ref{psin}) leads to 
\begin{equation}
\label{eq20}
\phi_0^{(n+1)}(r)=\int_0^{\infty} dr' K_0(r,r')\, \phi_0^{(n)}(r')\quad n \geq 1,
\end{equation}
with $\phi_0^{(0)}(r)=r \sqrt{v(r)}$ and $K_0(r,r')= \sqrt{v(r)} \inf(r,r') \sqrt{v(r')}$. Clearly we recover 
\begin{equation}
\label{eq22}
\phi_0^{(n+1)}={\cal K}_0\, \phi_0^{(n)}={\cal K}^{(n+1)}_0\,\phi_0^{(0)}\quad  n \geq 0, 
\end{equation}
in terms of the iterated kernel ${\cal K}^{(n)}_0$ (see section \ref{sec2.1}). As
\begin{equation}
\lim_{r \to \infty} \psi_n(r)=\int_0^{\infty} dr\, r\,  v(r) \, \psi_{n-1}(r) \quad n \geq 1, 
\nonumber
\end{equation}
is in fact the scalar product
\begin{equation}
\label{eq21}
\lim_{r\rightarrow \infty}\psi_{n}(r)= \langle \phi_0^{(0)}| \phi_0^{(n-1)}\rangle,
\end{equation}
we deduce that $\omega_{n}=\delta_{n-1}$ when the starting function in (\ref{delt}) is $\phi_0(r)=r \sqrt{v(r)}$. This shows that $\omega_n$ is an increasing sequence converging to $1/g^{(0)}_{\rm{c}}$.

Now, let us consider the sequence $\alpha_n$. We show in the \ref{app1} that $n \mapsto \alpha_n$ is a decreasing sequence which converges  to $1/g^{(0)}_{\rm{c}}$. In the \ref{app1} we relate all the $\psi'_n(0)$ entering the definition of $\alpha_n$ to the Jost function for $V(r)$ at zero energy. This is of some interest in the measure where we obtain lower limits for $g^{(0)}_{\rm{c}}$ circumventing the constraint $g^{(0)}_{\rm{c}} \int_0^{\infty} dr \, r |v(r)| < 2$ of \cite{la97}.

The sequence $\alpha_n$ and $\omega_n$ are simple enough to write the first members explicitly. First of all we remark that
\begin{equation}
\alpha_0=\int_0^{\infty}dx\, x\, v(x),
\end{equation}
which is just the Bargmann-Schwinger necessary condition for the existence of bound states. We also have $\omega_0=0$. The next order yields the following relation
\begin{equation}
\fl \frac{\int_0^{\infty}dx\, x\, v(x)}{\int_0^{\infty}dx\, v(x)\int_0^{\infty} dy\,\inf(x,y)\, y\, v(y)}=\frac{1}{\alpha_1}\leq g^{(0)}_{\rm{c}}\leq \frac{1}{\omega_1}=\frac{\int_0^{\infty}dx\, x^2\, v(x)}{2\int_0^{\infty}dx\, x\, v(x)\int_0^x dy\, y^2\, v(y)},
\end{equation}

\section{Tests of the bounds on some common potentials}
\label{sec4}

In this section, we propose to test the accuracy of the various upper and lower limits obtained in the sections \ref{sec2} and \ref{sec3} with three potentials: a square well potential (denotes hereafter SW potential), 
\begin{equation}
\label{eq3.1}
V(r)=-gR^{-2}\, \theta(1-r/R);
\end{equation}
an exponential potential (denotes hereafter E potential)
\begin{equation}
\label{eq3.2}
V(r)=-gR^{-2}\, \exp(-r/R);
\end{equation}
a non monotonic potential (denotes hereafter PE potential)
\begin{equation}
\label{eq3.3}
V(r)=-gR^{-3}\, r\exp(-r/R);
\end{equation}

In these potentials, the radius $R$ is arbitrary (but positive). Due to the scaling property, which does not affect the critical value $g_{\rm{c}}^{(\ell)}$, the radius $R$ appearing in the potential can be set to unity. 

In order to test the reliability of the different methods when the convexity of the potential increases, we perform simple analytical calculations (that we do not report here) involving the SW potential (\ref{eq3.1}), when the angular momentum $\ell$ increases. We also study the E and the PE potentials only for the S-wave. 

In order to examine in what extent our recursive procedure lead rapidly to the exact result, the exact value of the critical coupling constants of the potentials, $g_{\rm{c}}^{(\ell)}$, and its approximated upper and lower limits investigated in the sections \ref{sec2} and \ref{sec3} are depicted in the Tables  \ref{tab1} to  \ref{tab4}. More precisely, the Kellogg's coefficients $\gamma_n$ are shown in the Table \ref{tab1}, the Kolom\'y's coefficient are reported in the Table \ref{tab2}. The results obtained with the variational method and its iterated are given in Table \ref{tab3}. In Table \ref{tab4}, we report the results obtained with lower and upper limits discussed in section \ref{sec3}. As a further information, we give in Table \ref{tab5} the results for $\alpha_4$ and $\omega_4$ which are compared to those obtained with the formulas (\ref{eq1}), (\ref{eq2}), (\ref{eq3}) and (\ref{eq4}) for the three potentials (\ref{eq3.1})-(\ref{eq3.3}).

In all cases, few iterations are enough to obtain strong restrictions on the possible values of $g_{\rm{c}}^{(\ell)}$, especially for low value of the angular momentum.

\section{Conclusions}
\label{sec5}

In this article we have presented in section \ref{sec2} several iterative procedures yielding sequences of upper limits on the critical value, $g_{\rm{c}}^{(\ell)}$, of the coupling constant (strength), $g$, of the potential, $V(r)=-g v(r)$, for which a first $\ell$-wave bound state appears. In section \ref{sec3} we have obtained a method yielding both upper {\it and} lower limits on $g_{\rm{c}}^{(\ell)}$. All these sequences converge rather rapidly to the exact critical value as shown by the tests presented  in section \ref{sec4}. Due to the construction of the sequences, for example $\beta_n=1/\omega_{2 n}$, the convergence of the sequences obtained in section \ref{sec2} is faster than the ones of section \ref{sec3}. However results which are the closest to the exact values are obtained, with minimal numerical efforts, from the combination of variational and iterative methods.

Note that the results presented in this article, of similar accuracy than the results given by other methods, can always be improved when use is made of a supplementary iteration. The accuracy of the results that we obtain can be  measured by the difference $1/\omega_n -1/\alpha_n$.

\appendix

\section{Monotony of the sequence $\alpha_n$}
\label{app1}

In this appendix we  study the sequence $\alpha_n$ defined by (\ref{alphan}). As made for $\omega_n$, we express $\alpha_n$ in terms of a scalar product. We know that $\psi_{n+1}(r) \simeq r \int_0^{\infty} dr\, v(r) \, \psi_{n}(r), n \geq 1 $ at the vicinity of $r=0$, so that its first derivative is the scalar product
\begin{equation}
\psi'_{n+1}(0)= \langle q_0 | \phi_0^{(n)}\rangle 
\end{equation}
with $q_0(r)=\sqrt{v(r)}$ and $\phi_0^{(n)}(r)=\sqrt{v(r)}\psi_{n}(r)$. 

The proof of the monotony is made in two steps. First we show that the inverse of the Jost function at zero energy, given for example in Eqs. (13) and (14) of Ref. \cite{la97}, in terms of the coupling constant $g$ is simply the series 
\begin{equation}
\label{ap0}
\frac{1}{f_0(g,0)}=\sum_{n=0}^{\infty} \psi'_{n}(0) \, g^n
\end{equation}
which converges for $g < g^{(0)}_{\rm{c}}$. It can be easily verified by considering $\tilde{\psi}_n(r)$, defined by the recurrence relation (\ref{psin}) but where the first term $\tilde{\psi}_0(r)$ is equal to unity. 
The equation (\ref{psin}) becomes 
\begin{eqnarray}
\tilde{\psi}_n(r) &= & M_n -  \int_r^{\infty} dr' \, (r'-r) \, v(r) \  \tilde{\psi}_{n-1}(r')  \quad\quad n \geq 1 \nonumber\\
M_n & = & \int_0^{\infty} dr \, r \, v(r) \, \tilde{\psi}_{n-1}(r) \quad\quad n \geq 1.
\end{eqnarray}

Introducing $\tilde{\phi}_0^{(n)}(r)= \tilde{\psi}_n(r)\, \sqrt{v(r)}$, $\phi_0^{(n)}(r)= \psi_n(r) \,\sqrt{v(r)}$ and 
$q_0(r)=\sqrt{v(r)}$ we have, due to the symmetry of the scalar product,
\begin{eqnarray}
M_n &=&\langle  \psi_0^{(0)} | \tilde{\phi}_0^{(n-1)} \rangle=
\langle  \psi_0^{(0)} | {\cal K}_0^{(n-1)}\tilde{\phi}_0^{(0)} \rangle= 
\langle  \psi_0^{(0)} | {\cal K}_0^{(n-1)} q_0 \rangle = 
\langle q_0 | {\cal K}_0^{(n-1)} \psi_0^{(0)}\rangle \nonumber \\
&=& \langle q_0 | \psi_0^{(n-1)}\rangle=\int_0^{\infty} dr \, v(r) \, \psi_{n-1}(r) = \psi'_{n}(0).
\end{eqnarray}
This allows to define $M_n$ for $n=0$ and we have $M_0=1$.

On the other hand, it has been shown that the zero energy Jost function could be written as \cite{la97}
\begin{equation}
\label{ap1}
f_0(g,0)=\sum_{n=0}^{\infty} (-)^n a_n \, g^n,
\end{equation}
with
\begin{equation}
\fl a_n  =  \int_0^{\infty} dr_1\, r_1\, v(r_1)\int_{r_1}^{\infty} dr_2\, 
(r_2-r_1)v(r_2) \ldots  \int_{r_{n-1}}^{\infty} dr_n\, (r_n-r_{n-1})v(r_n) \quad
n \geq 2 
\label{ap2}
\end{equation}
and $a_0=1,a_1 =\int_0^{\infty} dr \ r v(r) $. We have the following relation between $M_n$ and $a_n$
\begin{equation}
\label{ap3}
\sum_{p=0}^n M_{n-p} \, a_{p} \, (-)^p =0 \quad\quad n \geq 1,
\end{equation}
whereas for $n=0$, $a_0 M_0=1$. This shows that the series $\sum_{n=0}^{\infty} M_n \, g^n$ is equal to $1/f(0,g)$ for $ g < g^{(0)}_{\rm{c}}$. Indeed, the relation (\ref{ap3}) is just the relation that exists between Taylor's coefficients of the 
functions $f(x)$ and $1/f(x)$. 

Secondly we prove the following lemma

{\bf lemma} {\it Let be the series 
\begin{equation}
F(g)=\sum_{n=0}^{\infty} \beta_n \, g^n 
\end{equation}
convergent for $g < R$, ($R >0$), such that $\beta_0=1$ and 
\begin{eqnarray}
(\forall n \geq 0) \quad\quad \beta_n & > & 0 \nonumber\\
(\forall n \geq 0) \quad\quad \beta_{n+2} \, \beta_n & \leq & \beta_{n+1}^2. 
\label{pro1}
\end{eqnarray}
Then, for every $\lambda >0 $, the series 
\begin{equation}
G(g)=\sum_{p=0}^{\infty} \beta_p \, g^p  \times \sum_{q=0}^{\infty} \lambda^q g^q=\sum_{n=0}^{\infty} \gamma_n \, g^n 
\end{equation}
convergent for $g < \tilde R=\inf(R,1/\lambda)$ is such that $\gamma_0=1$ and has the property  
\begin{eqnarray}
(\forall n \geq 0) \quad\quad \gamma_n & > & 0 \nonumber\\
(\forall n \geq 0)  \quad\quad \gamma_{n+2} \, \gamma_n & \leq & \gamma_{n+1}^2. 
\label{pro2}
\end{eqnarray}}

The proof of the lemma originates from the definition of $\gamma_n$
\begin{equation}
\gamma_n=\sum_{p=0}^n \beta_p \, \lambda^{n-p}
\end{equation}
which asserts that $\gamma_0=1$ and $\gamma_n >0$ since $\lambda$ and $ \beta_n$ are positive (see (\ref{pro1})). The second property in (\ref{pro2}) is satisfied if and only if the following relation is satisfied for $n \geq 0$
\begin{equation}
\sum_{p=0}^{n+2} \beta_p \, \lambda^{n+2-p} \, \sum_{q=0}^n \beta_q \ \lambda^{n-q} - \left( \sum_{p=0}^{n+1} \beta_p \, \lambda^{n+1-p} \right)^2  \leq 0. 
\label{ineq1}
\end{equation}
The inequality (\ref{ineq1}) is equivalent to the requirement that 
\begin{equation}
(\beta_{n+2}- \beta_{n+1} \lambda)  \sum_{p=0}^n \beta_p \, \lambda^{n-p} - \beta_{n+1}^2 \leq 0 ,
\end{equation}
which is manifestly satisfied when
\begin{equation}
(\forall n, 1 \leq p \leq n) \quad\quad  \beta_{n+2} \beta_p - \beta_{n+1}\beta_{p+1} \leq 0. 
\end{equation}
The relation (\ref{pro1}) and the positivity of $\beta_n$ imply that this latter inequality, which originates from the iterated of (\ref{pro1}) when $n$ is lowered up to $p$, is verified.

Now we use the fact that, for any $g < g_1= g^{(0)}_{\rm{c}}$ the Jost function can be written as \cite{newt}
\begin{equation}
f_0(g,0)=\prod_{n=1}^{\infty} \left( 1- \frac{g}{g_n} \right)
\label{prod}
\end{equation}
where the $g_n$'s still denote the characteristic numbers of the eigenvalue problem considered. The product (\ref{prod}) exists for $g < g_1$ when each series $\sum_{n=0}^{\infty} g_n^{-p} $ converges for every integer $p \geq 1$, which is true when the (positive) trace of the iterated kernel $K_0^{(n)}(r,r')$ is finite. The equation (\ref{prod}) implies that for any $g < g_1$ we have 
\begin{eqnarray}
\frac{1}{f_0(g,0)} & = &\prod_{n=1}^{\infty} \sum_{p=0}^{\infty} \left( \frac{g}{g_n} \right)^p = \lim_{N \to \infty} S_N \nonumber\\
 S_N & = & \prod_{n=1}^{N} \sum_{p=0}^{\infty} \left( \frac{g}{g_n} \right)^p .
\label{prod2}
\end{eqnarray}
Since all the quantities in (\ref{prod2}) are positive we can write, for $g < g_1$:
\begin{equation}
S_N = \sum_{p=0}^{\infty} s^{(N)}_p \, g^p
\end{equation}
where $s^{(N)}_0 \equiv 1$. Now assuming that for some $N$, the following property holds:
\begin{equation}
(\forall p \geq 0) \quad\quad s^{(N)}_{p+2} \, s^{(N)}_p - \left( s^{(N)}_{p+1} \right)^2  \leq 0 
\label{recc}
\end{equation}
according to the lemma, the property still holds for $N+1$ as well. Note that the radius of convergence $\tilde R$ of the lemma is always minorated by $g_1 >0$. For $N=1$, $s_p^{(1)}$ is simply $1/g_1^p$ and the property (\ref{recc}) is valid. 
For $N=2$, $s_p^{(2)}=\sum_{k=0}^{p} g_1^{-k}g_2^{k-p}$ and it can be verified that the property (\ref{recc}) is again valid. Therefore for every $N \geq 2$ the property (\ref{recc}) is also valid and in particular for $N$ going to infinity.
The comparison between (\ref{ap0}) and (\ref{prod2}) shows that $\psi'_n(0)=s_n^{(\infty)}$. From the relation (\ref{recc}) we obtain
\begin{equation}
(\forall n \geq 0) \quad\quad \frac{\psi'_{n+2}(0)}{\psi'_{n+1}(0)}  \leq  \frac{\psi'_{n+1}(0)}{\psi'_n(0)},
\end{equation}
which, with the definition (\ref{alphan}) of $\alpha_n$, prove that the sequence $\alpha_n$ is decreasing.

Since $\alpha_n$ is decreasing and positive it converges towards some $\alpha \geq 0$. On the other hand we know that the radius of convergence of the series $1/f(g,0)$ is $g_1$. Using the d'Alembert rule for the  series of positive numbers $(\psi'_n(0))_{n \geq 0}$ we have $1/g_1=\lim_{n \to \infty} \psi'_{n+1}(0)/\psi'_n(0)=\alpha$ so that $\alpha_n$ converges towards $1/g_1=1/g^{(0)}_{\rm{c}}$.

\ack

One of us (FB) would like to thank the FNRS for financial support (FNRS Postdoctoral Researcher position).

\newpage

\begin{table}
\protect\caption{Comparison between the coefficient $\gamma_n$ obtained with the Kellogg's method, with $\phi^{(0)}_0=r \sqrt{v(r)}$, and the exact results for the potentials (\protect\ref{eq3.1})-(\protect\ref{eq3.3}).} 
\label{tab1}
\begin{center}
\begin{tabular}{cccccc}
\br
Potential & $\gamma_1$ & $\gamma_2$ & $\gamma_3$ & $\gamma_4$ & Exact \\ 
\mr
E ($\ell=0$) & 1.5323 & 1.4480 & 1.4459 & 1.4458 & 1.4458 \\
PE ($\ell=0$) & 0.70463 & 0.67718 & 0.67669 & 0.67668 & 0.67668 \\
SW ($\ell=0$)  & 2.4853  & 2.4676 & 2.4674 & 2.4674 & 2.4674 \\
SW ($\ell=1$)  & 10.247  & 9.8885 & 9.8707 & 9.8697 & 9.8696 \\
SW ($\ell=2$)  & 21.635  & 20.317 & 20.202 & 20.192 & 20.191 \\
SW ($\ell=3$)  & 36.630  & 33.620 & 33.278 & 33.227 & 33.217 \\
SW ($\ell=4$)  & 55.210  & 49.745 & 49.000 & 48.864 & 48.831 \\
SW ($\ell=5$)  & 77.374  & 68.669 & 67.322 & 67.039 & 66.954 \\ 
\br
\end{tabular}
\end{center}
\end{table}

\begin{table}
\protect\caption{Comparison between the coefficient $\beta_n$ obtained with the Kolom\'y's method, with $\phi^{(0)}_0=r \sqrt{v(r)}$, and the exact results for the potentials (\protect\ref{eq3.1})-(\protect\ref{eq3.3}).} 
\label{tab2}
\begin{center}
\begin{tabular}{cccccc}
\br
Potential & $\beta_1$ & $\beta_2$ & $\beta_3$ & $\beta_4$ & Exact \\ 
\mr
E ($\ell=0$) & 1.4674 & 1.4465 & 1.44582 & 1.44580 & 1.4458 \\
PE ($\ell=0$) & 0.68270 & 0.67682 & 0.67669 & 0.67668 &  0.67668 \\
SW ($\ell=0$)  & 2.4706 & 2.46744 & 2.4674 & 2.4674 & 2.4674 \\
SW ($\ell=1$) & 10.000 & 9.8770 & 9.8701 & 9.8696 & 9.8696 \\
SW ($\ell=2$)  & 20.811  & 20.253 & 20.198 & 20.192 & 20.191 \\
SW ($\ell=3$)  & 34.851 & 33.439 & 33.252 & 33.223 & 33.217 \\
SW ($\ell=4$)  & 52.1053 & 49.372 & 48.935 & 48.852 & 48.831 \\
SW ($\ell=5$)  & 72.5672 & 68.022 & 67.192 & 67.039 & 66.954 \\ 
\br
\end{tabular}
\end{center}
\end{table}

\begin{table}
\protect\caption{Comparison between the upper limit on $g_{\rm{c}}^{(\ell)}$ obtain with the variational method, the combination of the variational and the Kellogg's method and the exact value of $g_{\rm{c}}^{(\ell)}$. The variational wave function $\phi_{\ell}(r)$ are $r \exp(-q r)$ for the E and the PE potentials and $r^p$ for the SW potential.}
\label{tab3}
\begin{center}
\begin{tabular}{ccccc}
\br
Potential & $ \phi_{\ell}$ & ${\cal K_{\ell}} \phi_{\ell}$ &  Exact \\ 
\mr
E ($\ell=0$)   & 1.44676 & 1.44582 & 1.44580 \\
PE ($\ell=0$)  & 0.68543 & 0.67672 &  0.67668 \\
SW ($\ell=0$)  & 2.4747 & 2.4674 & 2.4674 \\
SW ($\ell=1$)  & 9.9934 & 9.8710 & 9.8696 \\
SW ($\ell=2$)  & 20.604 & 20.201 & 20.191 \\
SW ($\ell=3$)  & 34.099 & 33.253 & 33.217 \\
SW ($\ell=4$)  & 50.357 & 48.915 & 48.831 \\
SW ($\ell=5$)  & 69.295 & 67.117 & 66.954 \\ 
\br
\end{tabular}
\end{center}
\end{table}

\begin{table}
\protect\caption{Comparison between the coefficient $\alpha_n$ and $\omega_n$ obtained with the new iterative method and the exact results for the potentials (\protect\ref{eq3.1})-(\protect\ref{eq3.3}).} 
\label{tab4}
\begin{center}
\begin{tabular}{cccccccccc}
\br
Potential & $\alpha_1^{-1}$ & $\alpha_2^{-1}$ & $\alpha_3^{-1}$ & $\alpha_4^{-1}$ & Exact & $\omega_4^{-1}$ & $\omega_3^{-1}$ & $\omega_2^{-1}$ & $\omega_1^{-1}$\\ 
\mr
E ($\ell=0$) & 1.3333 & 1.4211 & 1.4408 & 1.4448 & 1.4458 & 1.4465 & 1.4495 & 1.4674 & 1.6000 \\
PE ($\ell=0$)  &0.64000 & 0.67009 & 0.67558 & 0.67650 & 0.67668 & 0.67682 & 0.67755 & 0.68270 & 0.72727 \\ 
SW ($\ell=0$) & 2.4000 & 2.4590 & 2.4664 & 2.4673 & 2.4674 & 2.4674 & 2.4677 & 2.4706 & 2.5000 \\
SW ($\ell=1$) & 8.5714 & 9.4839 & 9.7638 & 9.8419 & 9.8696 & 9.8770 & 9.9000 & 10.000 & 10.500 \\
SW ($\ell=2$) & 15.556 & 18.271 & 19.452 & 19.921 & 20.191 & 20.252 & 20.381 & 20.811 & 22.500 \\
SW ($\ell=3$) & 22.909 & 28.077 & 30.813 & 32.149 & 33.217 & 33.439 & 33.801 & 34.851 & 38.500 \\
SW ($\ell=4$) & 30.462 & 38.506 & 43.336 & 46.041 & 48.831 & 49.372 & 50.121 & 52.105 & 58.500 \\
SW ($\ell=5$) & 38.133 & 49.336 & 56.671 & 61.201 & 66.954 & 68.022 & 69.322 & 72.567 & 82.500 \\
\hline
\end{tabular}
\end{center}
\end{table}

\begin{table}
\protect\caption{Comparison between previously known upper and lower limits (\protect\ref{eq1}), (\protect\ref{eq2}), (\protect\ref{eq3}) and (\protect\ref{eq4}) on $g_{\rm{c}}^{(\ell)}$ and the exact results for the potentials (\protect\ref{eq3.1})-(\protect\ref{eq3.3}).} 
\label{tab5}
\begin{center}
\begin{tabular}{cccccccc}
\br
Potential & Eq. (\protect\ref{eq1})&Eq. (\protect\ref{eq2})&Eq. (\protect\ref{eq3})&Eq. (\protect\ref{eq4}) & Exact 
& $\alpha_4^{-1}$ & $\omega_4^{-1}$\\ 
\mr
E ($\ell=0$) & 1.4383 & 1.6755 & 1.5442 & 1.4467 & 1.4458       & 1.4448  & 1.4465  \\
PE ($\ell=0$) & 0.67421 & 0.76498 & 0.86547 & 0.67691 & 0.67668 & 0.67650 & 0.67682 \\
SW ($\ell=0$) & 2.3593 & 2.6667 & 4.0000 & 2.4747  & 2.4674     & 2.4673  & 2.4674  \\
SW ($\ell=1$) & 9.1220 & 11.719 & 10.068 & 9.9934  & 9.8696     & 9.8419  & 9.8770  \\
SW ($\ell=2$) & 18.454 & 25.413 & 20.895 & 20.604  & 20.191     & 19.921  & 20.252  \\
SW ($\ell=3$) & 30.245 & 43.570 & 35.424 & 34.099  & 33.217     & 32.149  & 33.439  \\
SW ($\ell=4$) & 44.425 & 66.089 & 53.519 & 50.357  & 48.831     & 46.041  & 49.372  \\
SW ($\ell=5$) & 60.947 & 92.909 & 75.114 & 69.295  & 66.954     & 61.201  & 68.022  \\
\br
\end{tabular}
\end{center}
\end{table}

\end{document}